\documentclass[sigconf,nonacm]{acmart}

\usepackage{enumitem}
\usepackage{mathtools}
\usepackage[disable]{todonotes}
\usepackage{todonotes}
\usepackage{amsthm}
\usepackage{algorithm}
\usepackage[noend]{algpseudocode}

\usepackage{commath}
\usepackage{comment}
\usepackage{multirow}
\usepackage{pgfplots, pgfplotstable, booktabs, colortbl, siunitx, array}
\usepackage{longtable}
\usepackage{subcaption}

\newcolumntype{R}[1]{>{\raggedleft\arraybackslash }b{#1}}

\newcommand{\reals}{\mathbb{R}}

\AtBeginDocument{%
  \providecommand\BibTeX{{%
    \normalfont B\kern-0.5em{\scshape i\kern-0.25em b}\kern-0.8em\TeX}}}

\begin{document}

\title{Trading via Selective Classification}


\author{Nestoras Chalkidis}
\affiliation{%
  \institution{Department of Computer Science}
  \country{University of Liverpool}
}
\email{n.chalkidis@liverpool.ac.uk}

\author{Rahul Savani}
\affiliation{%
	\institution{Department of Computer Science}
  \country{University of Liverpool}
}
\email{rahul.savani@liverpool.ac.uk}

\begin{abstract}

A binary classifier that tries to predict if the price of an asset 
will increase or decrease naturally gives rise to a
trading strategy that follows the prediction and thus always has a position in
the market.
Selective classification extends a binary or many-class classifier to allow
it to abstain from making a prediction for certain inputs, thereby allowing a
trade-off between the accuracy of the resulting selective classifier
against coverage of the input feature space.
Selective classifiers give rise to trading strategies that do not take a
trading position when the classifier abstains.
We investigate the application of binary and ternary selective
classification to trading strategy design.
For ternary classification, in addition to classes for the price going up or
down, we include a third class that corresponds to relatively small price moves
in either direction, and gives the classifier another way
to avoid making a directional prediction.
We use a walk-forward train-validate-test approach to evaluate and compare
binary and ternary, selective and non-selective classifiers across several
different feature sets based on four classification approaches: logistic
regression, random forests, feed-forward, and recurrent neural networks. 
We then turn these classifiers into trading strategies for which we perform
backtests on commodity futures markets.
Our empirical results demonstrate the potential of selective classification for
trading.
	
\end{abstract}



\keywords{time series prediction, selective classification, trading strategy}

\maketitle

\section{Introduction}

This paper studies the fundamental and well-studied problem of financial price
time series prediction. 
Specifically, we apply binary and ternary machine learning (ML) classifiers to
intraday futures time series to predict if the next period's price will increase
or decrease, with a third class in the ternary case that corresponds to relatively
small price moves in either direction.
The novelty of our work is to apply \emph{selective classification} (also known 
as classification with a reject option, and classification with abstention), which
allows a trained classifier to abstain from making a prediction.

We turn the selective and non-selective classifiers into
trading strategies that take a position for the next period based on the 
classifier's prediction when it makes one.
The selective classifiers are able to not take a position if the classifier
is not suitably confident about its prediction.
We perform cross-validated backtests using a walk-forward approach for the
resulting trading strategies and analyse the results, which show the promise 
of selective classification for trading strategy design.
To the best of our knowledge, the application of selective
classification to trading strategy design has not been explored in the literature.

Our key contributions are the following:
\begin{itemize}[leftmargin=0.3cm]
\itemsep1mm
\item We train, evaluate, and compare binary and ternary selective classifiers
	using four different ML classification approaches
	(Section~\ref{sec:classification}): logistic regression, random
	forests, feed-forward networks, and Long-Short Term
	Memory (LSTM) networks.
 
\item We compare selective and non-selective classifiers in terms of their
	accuracy. We present the accuracy coverage trade-off of the selective
	classifiers, and show that they have better accuracy 
	compared to the non-selective classifiers (Section~\ref{sec:classresults}). 


\item We perform walk-forward cross-validated backtests of trading strategies
	based on the classifier's predictions (Section \ref{sec:trading}). We find that
	the selective classifiers give better risk-adjusted performance, with several models
	remaining profitable even with a reasonable level of slippage included,
	which shows the potential of selective classification for trading strategy design
	(Section~\ref{sec:trading_results}).
	 
\end{itemize}

\section{Related Work} 

\noindent
\emph{Financial time series forecasting.} A long line of work has studied 
predictability of financial time series. In recent years, with the rise of 
Deep Learning (DL), much of the focus of work in this area has focussed on DL.
A recent survey on financial time
series forecasting using ML, and in particular DL, is provided
in~\cite{SGO20}. 
The survey shows a comprehensive review of many
studies that apply ML and DL for financial time series forecasting. 
Most of the studies were focused on stock price
forecasting and the most commonly used models were LSTM networks~\cite{hochreiter1997long}.
Here we present a non-exhaustive but indicative selection of examples.
Shihao Gu et. al.~\cite{GKX20} present a comparison of different ML methods for
measuring risk premia in empirical asset pricing. 
\cite{bao2017deep} present a DL framework that uses a wavelet
transform to de-noise stock price time series and stacked auto-encoders to
produce high-level features. 
\cite{kim2019forecasting} develop a DL approach using a so-called feature fusion long short-term
memory-convolutional neural network (LSTM-CNN). 
\cite{patel2015predicting} compare artificial neural networks,
support vector machines, random forest, and naive-Bayes for prediction of 
stock price movements in Indian stock markets.
\cite{borovkova2019ensemble} applied an ensemble of LSTM networks to predict 
intraday stock prices with technical indicators as input features.

\smallskip

\noindent
\emph{Selective classification.} The concept of abstention or rejection in
classification has a long history. It was introduced in 1970~\cite{Chow70}. In
the same year, Hellman investigated $(k, k')$ nearest neighbors
with a rejection rule~\cite{Hellman70}.
Much more recently,~\cite{BartlettW08} considered
binary classification where the classifier can abstain from making a prediction
but then incurs a cost. 
In~\cite{CortesDM16}, a boosting algorithm for binary classification with
abstention was presented for the same case where abstention has a cost.

The term ``selective classification'' was introduced in~\cite{El-YanivW10},
which studied the risk-coverage trade-off, and constructs algorithms that
near-optimally achieve the best trade-off, where in contrast to earlier models
there is no direct cost for abstention.
\cite{WienerE15} extend the results ~\cite{El-YanivW10} to
the noisy and agnostic setting (where almost not assumptions are made about the
best model).

Given a trained neural network, the authors of~\cite{GeifmanE17} proposed a
method to construct a selective classifier. At test time, the classifier
rejected instances as needed to grant the desired risk with high probability.
The proposed classification mechanism was based on applying a
selected threshold on the maximum neuronal response of the softmax layer. The
results indicate that even for challenging data sets selective classifiers
are extremely effective, and with appropriate coverage 
surpassed the then-best-known results on ImageNet. 
Our study is based on~\cite{GeifmanE17}.

In~\cite{GeifmanE19} the authors considered the problem of selective
classification in deep neural networks, by developing an architecture with an
integrated rejection option (SelectiveNet). Their goal was to simultaneously optimize during
training both classification and rejection (in contrast to~\cite{GeifmanE17} which
assumes a pre-trained classifier). 

In~\cite{ThulasidasanBBC19}, a method to combat label noise when training
deep neural networks for classification was proposed. A loss function was used
which permitted abstention during training thereby allowing the deep neural
networks to abstain on confusing samples while continuing to learn and improve
classification performance on the non-abstained samples.


\section{Technical preliminaries} 

Here we introduce the ``Selection with Guaranteed Risk'' method for selective 
classification from~\cite{GeifmanE17} that we use in this paper.
Our exposition follows closely that of~\cite{GeifmanE17}.

For a multi-class classification problem, let $X \subseteq \reals^k$ for $k$
real-valued features denote the feature space, and $Y=\left\{1, 2,\dots,
k\right\}$ the finite set of $k$ classes (labels).
Let $P(X, Y)$ be the underlying, unknown distribution over $X\times Y$. 
A classifier is defined as $f:X\rightarrow Y$, and the true risk of this
classifier w.r.t $P$ is given by $R(f|P)\triangleq E_{P(X,Y)}\left[\ell(f(x),
y)\right]$, where $\ell:Y\times Y\rightarrow \!R^{+}$ is a loss function.
%
The empirical risk of a classifier $f$ given a training set $S_{m}=\{(x_i,
y_i)\}_{i=1}^{m}\subseteq (\textit{X}\times \textit{Y})^m$ sampled i.i.d from 
$P(X, Y)$ is defined as
$\hat{r}(f|S_m)\triangleq \frac{1}{m}\sum_{i=1}^{m}\ell(f(x_i), y_i)\text{.}$

A selective classifier~\cite{El-YanivW10} is a pair of $(f, g)$
functions, where $f$ is a classifier and $g$ is a 
selection function, $g: \textit{X}\rightarrow \{0, 1\}$:
\begin{equation}\label{selection function}
(f, g)(x)\triangleq 
	\begin{dcases}
    f(x),& \text{if } g(x)=1\text{;}\\
    \text{don't know}, & \text{if } g(x)=0\text{.}
	\end{dcases}
\end{equation}
The selective classifier abstains iff $g(x)=0$, otherwise the
prediction of the classifier is given by~$f$. 
 
The performance of the selective classifier is measured according to its
coverage and selective risk. 
The coverage is $\phi(f, g)\triangleq E_{P}\left[g(x)\right]$, and
represents the expectation under $P$ of the number of the non-rejected samples.
The selective risk is defined as:
\begin{equation}\label{selective risk}
R(f, g)\triangleq \frac{E_{P}\left[\ell(f(x), y)g(x)\right]}{\phi(f, g)}\text{.}
\end{equation}
According to~\eqref{selective risk}, the risk of a selective
classifier can be traded-off for coverage:
Given a classifier $f$, training set $S_{m}$, confidence parameter
$\delta>0$, and a desired risk target $r^{*}>0$, the goal is to use $S_m$ to 
create a selection function $g$ such that the selective risk of $(f,g)$ satisfies:
\begin{equation}\label{condition}
\textbf{Pr}_{S_{m}}\{R(f, g) > r^{*}\} < \delta\text{,}
\end{equation}
where the probability is over training sets, $S_{m}$, sampled i.i.d.
from the unknown underlying distribution $P$. 
Among those that satisfy~\eqref{condition}, the best classifiers are those that
maximize coverage.

For $\theta >0$, the selection function $g_{\theta}: \textit{X}\rightarrow \{0, 1\}$
is defined as:
\begin{equation}\label{selection function2}
g_{\theta}(x) = g_{\theta}(x|\kappa_f)\triangleq  
	\begin{dcases}
    1,& \text{if } k_f(x)\geq \theta \text{;}\\
    0, & \text{otherwise}\text{,}
	\end{dcases}
\end{equation}
where $\kappa_f$ is a confidence rate function $\kappa_f: \textit{X}\rightarrow
\!R^{+}$ for $f$.

After defining the selection function, the empirical selective risk of any
selective classifier $(f, g)$ given a training set $S_{m}$ is given by:
\begin{equation*}
\hat{r}(f, g|S_m)\triangleq \frac{\frac{1}{m}\sum_{i=1}^{m}\ell(f(x_i), y_i)g(x_i)}{\hat{\phi}(f, g|S_m)}\text{,}
\end{equation*}
where $\hat{\phi}$ is the empirical coverage, $\hat{\phi}(f, g|S_m)\triangleq
\frac{1}{m}\sum_{i=1}^{m}g(x_i)$. The $g$ projection of $S_m$ is
$g(S_m)\triangleq \{(x, y)\in S_m: g(x)=1\}$. 

In Algorithm~\ref{SGR}, the Selection with Guaranteed Risk (SGR) algorithm
from~\cite{GeifmanE17} is presented. The algorithm finds the optimal bound
guaranteeing the required risk
with sufficient confidence by applying a binary search. The SGR outputs a risk
bound $b^{*}$ and a selective classifier $(f, g)$. 
Lemma~3.1 in~\cite{GeifmanE17},
gives the tightest possible numerical bound generalization for a single
classifier based on a test over a labelled sample.

\begin{algorithm}
\caption{Selection with Guaranteed Risk (SGR)}
\label{SGR}

\begin{algorithmic}[1]

\State SGR$(f, k_f, \delta, r^*, S_m)$
\State Sort $S_m$ according to $k_{f}(x_i)$, $x_i\in S_m$ (and now assume w.l.o.g. that indices reflect this ordering).
\State $z_{min}=1$; $z_{max}=m$
\For{$i=1$ \textbf{to} $k\triangleq \lceil log_{2}m \rceil$}
	  \State $z=\lceil (z_{min} + z_{max})/2 \rceil$
	  \State $\theta = k_{f}(x_z)$
	  \State $g_i = g_{\theta}$ \{see~\eqref{selection function2}\}
	  \State $\hat{r}_i = \hat{r}(f, g_{i}|S_m)$
	  \State $b_i^* = B^*(\hat{r}_i, \delta/\lceil log_{2}m \rceil, g_{i}(S_m))$ \hfill see Lemma~3.1 in \cite{GeifmanE17}
      
	\If {$b_i^* < r^*$}
		\State $z_{max} = z$
	\Else
		\State $z_{min} = z$
	\EndIf
	\State \textbf{end if}

\EndFor
\State \textbf{end for}

\State Output - $(f, g_{k})$ and the bound $b_k^*$.
\end{algorithmic}
\end{algorithm}

\noindent
In the rest of the paper, all selective classifiers are built
using~Algorithm~\ref{SGR}. All classification methods we use output class
probabilities; we use the maximum class probability for $k_f$, and we use
$\delta = 0.001$ (both in line with~\cite{GeifmanE17}).






\section{Classification}
\label{sec:classification}

\subsection{Classification methodology}

\subsubsection{Raw data}

In this study, we used data from five metal commodities futures markets,
specifically, Gold (GC), Copper (HG), Palladium (PA), Platinum (PL) and Silver
(SI), as traded on the Chicago Mercantile Exchange's (CME) Globex electronic
trading platform. 
These futures markets trade 24 hours a day with a 60-minute break each day at
5:00pm (4:00 p.m. CT). The raw data corresponds to the time period 14-02-2011
00:00 to 31-05-2019 17:00, which corresponds to 97482 30-minute intervals. 


\subsubsection{Data preprocessing - labelling the data}

In the supervised binary classification problem, each sample has a corresponding
label which is defined based on the closing logarithmic return price
(clrp) of that sample\footnote{For the binary case, the label could equivalently
be defined just with the change in close price, but since we actually
use the return value for the ternary case, we use it also here.}. The
label of each sample is defined as follows:

\begin{equation*}
\text{label} =
\begin{cases}
	-1 \text{ or short}, & \text{if } clrp\leq0\ ,\\
	\phantom{-}1 \text{ or long}, & \text{if } clrp>0\ .
\end{cases}
\end{equation*}

In real-world trading strategies one has the option to be flat and not hold
a position in a given security.
Motivated by this, in addition to binary classification, we 
also explore ternary classification problem, where one more label corresponding
to ``flat'' is added. 
Intuitively, this third class will be defined to correspond to price moves that are 
small in absolute value, and to that end
we define a threshold value as follows.
The threshold value is defined as the product of the rolling volatility of the
closing simple return price over the last $k$ days with a multiplier value, and
compared with the volatility of the closing simple return price (csrpv). 
We set $k=48*30$, which represents the previous one month data, and we use four
different multiplier values of $\{0.3, 0.6, 0.9, 1.2\}$. 
Thus for the supervised ternary classification problem we have
four different cases where we alter the class distributions based on the
aforementioned multiplier values. Lower multiplier values indicate that there
are more -1 and 1 labels and less 0 labels.
By increasing the multiplier's value the number of 0 labels increases
and the number of -1 and 1 labels decreases.

\begin{equation*}
\text{label} =
		\begin{cases}
			-1 \text{ or short}, & \text{if } \quad \ csrpv \ < -\text{threshold},\\
			\phantom{-}0 \text{ or flat}, & \text{if } \quad \abs{csrpv} \leq \phantom{-}\text{threshold},\\
			\phantom{-}1 \text{ or long}, & \text{if } \quad \ csrpv \ > \phantom{-}\text{threshold}.
		\end{cases}
\end{equation*}

The classes are particularly imbalanced when the multiplier's value equals $0.3$ or
$1.2$. 
Standard techniques for imbalanced classes are under-sampling, over-sampling,
and class weighting.
For our sequential time series data, under-sampling and over-sampling are
problematic, so we used class weighting where the loss assigns more weight to
data from the under-represented classes.

\subsubsection{Data preprocessing - feature construction}

In our empirical study, a range of different features and combinations of them
were used as inputs to the four ML classifiers as there is an interest to investigate
their effect to the learning process. In particular, four different, expanding
feature sets, were investigated.
We call these feature sets FS1, FS2, FS3, and FS4, and our base set of features,
FS1 is contained in the remaining three feature sets, and so on:
$$\text{FS1} \subset \text{FS2} \subset \text{FS3} \subset \text{FS4}.$$
There are two reasons for this setup.
Firstly, by growing larger, richer feature sets in this way, we explore whether
the classifiers that we train are able to learn from and exploit the added
features.
Secondly, by considering four different feature sets, we are able to investigate
the benefits of binary versus ternary, and selective versus non-selective
classification within different settings, thereby seeing if a consistent picture
emerges of which approach appears better (indeed in both our classification and
backtesting results, we do see consistent insights across these feature sets).

\medskip

\noindent
\textbf{Basic price and volume features.}
To construct our basic feature set, FS1, we first constructed standard
price-volume time bars that comprise the Open, High, Low, Close, and Volume
(OHLCV) associated with 30-minute periods. 
We created these time-volume bars from raw tick data.
Then, since prices are typically highly non-stationary\footnote{
A stationary time series is one whose statistical properties, such as its mean
and standard deviation, do not depend on the time of an observation, i.e., they
are constant.}, we used logarithmic returns of the OHLC features, which are more
likely to be stationary.

We preprocessed trading volume independently.
Firstly, we observed that the trading volume appears to follow a power law distribution and is
positively skewed. 
We tested several popular methods for normalizing positively skewed data and in
the end chose the Box-Cox transformation~\cite{box1964analysis} as our method in
order to transform trading volume into a (more) stationary feature. 

Finally, for the logarithmic returns of the OHLC, and the Box-Cox normalised
volume, we applied a temporal normalization.
This contextualises the features according to recent past (it is of course crucial
to not use future data for this normalization in order to avoid lookahead bias).
This ``min-max'' normalization used a rolling window and produces scaled features
in $[0, 1]$ as follows:
\begin{equation*}
\displaystyle\hat{x} = \frac{x-rolling\_min(x)}{rolling\_max(x) - rolling\_min(x)}\text{,}
\end{equation*}
where $x$ is the value of a sample of a specific feature at a specific time, and
$\hat{x}$ is the normalized value of sample $x$. 
Our rolling window for the normalization corresponded to 10 days of data (which
was not optimized).
The ``min-max'' normalized OHLCV features comprise FS1, our most basic feature set.
FS1 contains 5 features.

\medskip

\noindent
\textbf{Moving average of return.}
Moving averages are very commonly used in ``technical analysis'' and trading
strategy design, as a natural way to summarize prices or volumes over different
time periods.
Moving averages act as low-pass filters, smoothing the signal and removing noise,
and therefore they are often used to identify the direction of trends.
To construct FS2, we added simple moving averages of the (min-max normalized)
close price feature from FS1 with three different lookbacks (window sizes).
The lookbacks that we used correspond to 1 day (48 30-minutes bars), 5 days
(240 30-minute bars), and 10 days (480 30-minute bars).
Thus, we add 3 new features to FS1 to form FS2, one for each of the three lookbacks.

\medskip

\begin{figure}[!h]
	\includegraphics[width=0.45\textwidth]{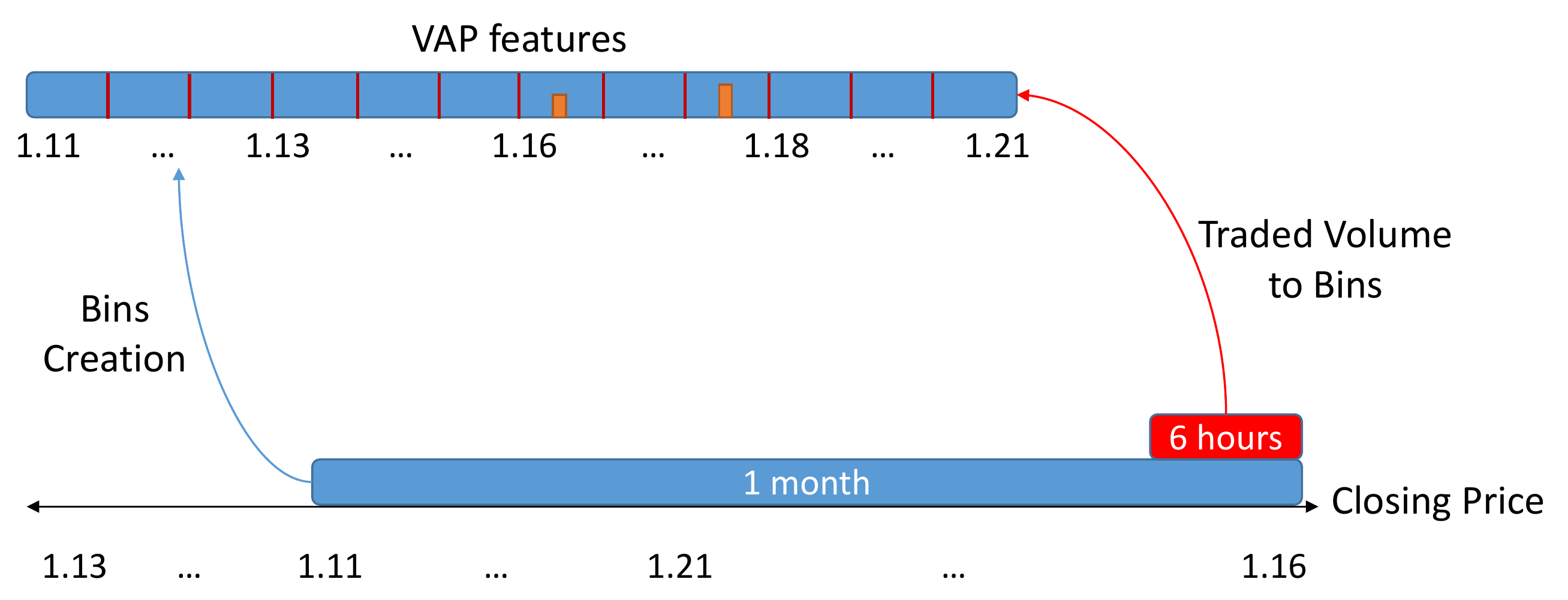}
    \caption{Illustration of VAP feature generation.}
    \label{VAP}
\end{figure}

\noindent
\textbf{Volume at price features.}
To enrich the feature set further, we consider features based on \emph{volume at
price (VAP)} (sometimes called ``Market Profile''~\cite{steidlmayer03}).
In essence, VAP analysis considers the histogram of traded volume at different prices
or price bins.
We created VAP-related features as follows, using two different windows as shown 
in Figure~\ref{VAP}.
One long-timeframe window defines a ``price context''; we use one month of 
data for this window.
The close price range spanned over this window is split into twelve equal size
bins.
We will have one feature for each of these bins.
The second window represents recent price action, in particular for the previous 
six hours of data.
Within this window, all traded volume is associated with the bin of the close 
price of the respective price-volume time bar.
Thus, we associate with each bin the total amount of volume traded at the prices
associated with the bin (as measured by close prices of the time bars) over the 
last six hours.
Finally, we normalize across the bins (divide by the total amount of volume over 
the last six hours), so that the corresponding twelve non-negative features 
sum to one and represent a distribution of volume (of the last six hours) within
the context of the price range over the last month.
FS3 is thus formed from FS2 by adding twelve additional features that take the 
form of a discrete probability distribution.

\medskip

\noindent
\textbf{Trade direction and aggressiveness features.}
Finally, we enrich the feature space with information that is derived from the
raw tick data and is intended to capture the aggressiveness of trading volume.
We use a method from~\cite{bouchaud2018trades} that classifies each trade (and
its associated volume) as ``aggressive'' or ``non-aggressive''.
If a trade triggers a price change (in practice because, for example, the volume
of an incoming market order is higher
than the available volume at the best quote on the opposite side of the limit order 
book) then the trade and its volume is marked as aggressive; otherwise
the trade and its volume is marked as non-aggressive.
\cite{bouchaud2018trades} found this type of classification had predictive value.

Using the raw tick data, we used this method to classify all trades as aggressive
or non-aggressive.
We further classify trades and the corresponding volume as \emph{buyer or seller
initiated} in the spirit of the Lee-Ready indicator~\cite{lee1991inferring}.
In a limit order book market, such as the futures market in our investigation, 
a trade is triggered by an incoming order, which is matched against sitting
order(s) that are already in the book.
If this incoming order is a buy (sell), which is normally apparent from the data
if one knows the best bid and ask when the order arrives\footnote{
When the tick data is ambiguous, standard rules such as using the last-used trade
direction, are applied to ensure that all trades/volume is classified as
buy/sell volume.}, then we classify this trade as a buy (sell) trade and volume.

 
With these two categorizations of trades and their volume into buy/sell and
aggressive/non-aggressive, we construct for a given 30-minute bar, the following
features (all min-max normalized in the same way as the FS1 features): 
the total number of trades 
(the total volume is already included in FS1), 
the difference between buyer and seller initiated trade count,
the difference between buyer and seller initiated volume, 
the non-aggressive volume,
the non-aggressive trade count,
the difference between buyer and seller initiated non-aggressive trade count, and
the difference between buyer and seller initiated non-aggressive volume (given that
we include total volume and trade count, we do not also include separate features
for aggressive volume/trade count, since the totals and the non-aggressive features
imply these).
FS4 is formed from FS3 by augmenting it with these seven features.

\subsubsection{Walk-forward cross-validation}
\begin{figure}[!h]
	\includegraphics[width=0.4\textwidth]{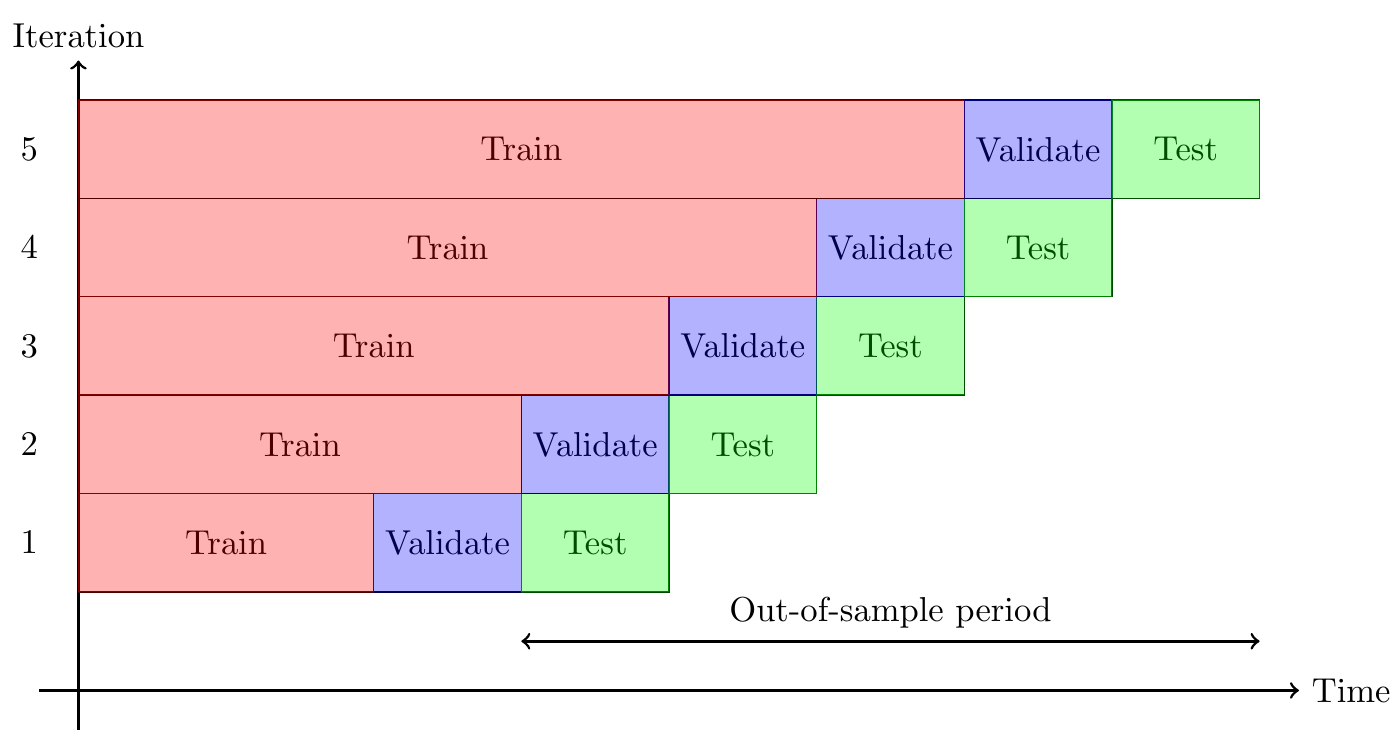}
    \caption{Walk-forward train-validation-test scheme.}
    \label{WF}
\end{figure}

Figure~\ref{WF} illustrates the anchored walk-forward scheme that was used to 
create train, validation and test sets.
The length of the initial train set used was 6 months; with the anchored scheme,
the length of each subsequent train set gets longer.
The length of all validation test sets used were 2 and 6 months respectively (a
shorter validation set was used so as to keep as more data for training and
testing).

\subsubsection{Hyperparameter Tuning}

As the design of the trading strategy will be based on the classifier's
predictions, one of our goals was to determine the best set of hyperparameters
that will result to the most efficient ML classifiers. Popular methods that are
used in the literature for hyperparameter tuning are the Grid Search, Random
Search and Bayesian optimization process. In this study, the Grid Search method
was used due to its simplicity. To make a fair comparison across all
classifiers, 12 hyperparameter combinations were used to define the best
hyperparameter set of each classifier. 

In both classification problems, at each walk-forward period, the best
hyperparameter set was selected based on the highest validation Matthews
Correlation Coefficient (MCC) value. 
%

%
We used MCC as our metric to define best models because it is
a balanced measure that takes into account true/false positives and negatives
and it can be used for both binary and ternary classification problems. 
Its advantages can be found in~\cite{chicco2020advantages}.

For logistic regression, we set the maximum number of iterations
equal to $\{250, 500\}$, the optimization algorithm to $\{liblinear, saga\}$, and
the inverse of regularization strength to $\{0.01, 0.001, 0.0001\}$. For all the
other hyperparameters, we used the default values from scikit-learn. 
For random forests~\cite{breiman2001random}, we used $\{500, 1000, 2000\}$ many
trees, splitting criteria in $\{gini, entropy\}$, and $sqrt(\#\text{features})$
and $log2(\#\text{features})$ for the number of features to consider when looking
for the best split.
For all other hyperparameters for random forests, default values in scikit-learn
were used. 

The feed-forward and LSTM networks consist from the same hyperparameter
combinations as they share a lot of common aspects. It is well known that neural
networks have too many hyperparameters to set. In this study, four different
network architectures and three different learning rates were defined. An
architecture is defined by (number of features, number of hidden units in the
first layer, number of hidden units in the second layer (if applicable), number
of units in the output layer depending on the classification labels). 
For both binary and ternary classification, the following architectures were used:
$(\#\text{features}, 512, \{2, 3\})$, 
$(\#\text{features}, 256, \{2, 3\})$, 
$(\#\text{features}, 512, 256, \{2, 3\})$, 
$(\#\text{features}, 256, 128, \{2, 3\})$. 
Learning values of $\{0.01, 0.001, 0.0001\}$ were used. 

The feed-forward networks consist of an input layer, one or more hidden layers
and an output layer. 
Rectified linear units~\cite{nair2010rectified} were used as activation
functions for all layers except the output layer, where softmax activation was
used. 
In hidden layers, $l_2$ weight regularization method was applied to prevent
overfitting.  
Batch normalization~\cite{ioffe2015batch} was used on each layer, where for each
batch it standardizes the inputs to a layer and reduces the number of training
epochs. 
Dropout~\cite{srivastava2014dropout} was also used as a further protection against
overfitting. 
The Adam optimizer with a decay of $1e^{-6}$ was used when the learning rate was
equal to $0.0001$; for the other learning rates, a decay of $1e^{-4}$ was used.

There are two different types of LSTM networks: stateless and stateful.
Stateless LSTM networks initialise the hidden and cell states freshly for each batch. 
Stateful LSTM networks pass to the second batch the hidden and cell states from
the first batch, and so on. 
In this study, we use the stateful LSTM networks as we want the long-term memory
to remember the content of the previous batches. 
In the LSTM layers, the hyperbolic tangent function was used as activation
function and the softmax activation function in the output layer. 
In the hidden layers the l2 weight regularization method was applied. Batch
normalization and dropout were also used. 
The Adam optimizer was used in the same way as on feed-forward networks. 
In the LSTM networks sequences of 48 (one day's) timesteps were used. 

\subsubsection{Selective Classification}

To create the selective classifiers, the SGR (Algorithm~\ref{SGR}) was applied on the
predicted probabilities for each non-selective binary/ternary classifier. 
Coverage is defined as the percentage of samples that have not been rejected by
the SGR algorithm (coverage for the non-selective classifiers is always 100\%). 
For different desired risk levels, the SGR algorithm applies selected thresholds
allowing a trade-off between accuracy and coverage. 
The selection of the best selective classifier threshold for both
binary and ternary was chosen to be the one that gives the highest MCC value (across
all 2 or 3 classes respectively).
Finally, for the ternary classification problem, the multiplier that 
determines which price moves get label 0, is selected based on the highest MCC
value of just the buy and sell labels, a choice driven by our goal of
having accurate buy/sell predictions as a basis for trading strategies.

\subsection{Classification results}
\label{sec:classresults}

Ultimately, we use the (selective and non-selective) classifiers that we train
as the basis for trading strategies.
Before we do that, in this section we first analyze the trained
classifiers, both selective and non-selective, purely on the classification task.
We explore the relative performance of selective versus non-selective
classification, binary versus ternary classification, the different
classification algorithms and the different feature sets.

\begin{figure}[!ht]
     \resizebox{\columnwidth}{!}{
     \includegraphics[width=\textwidth]{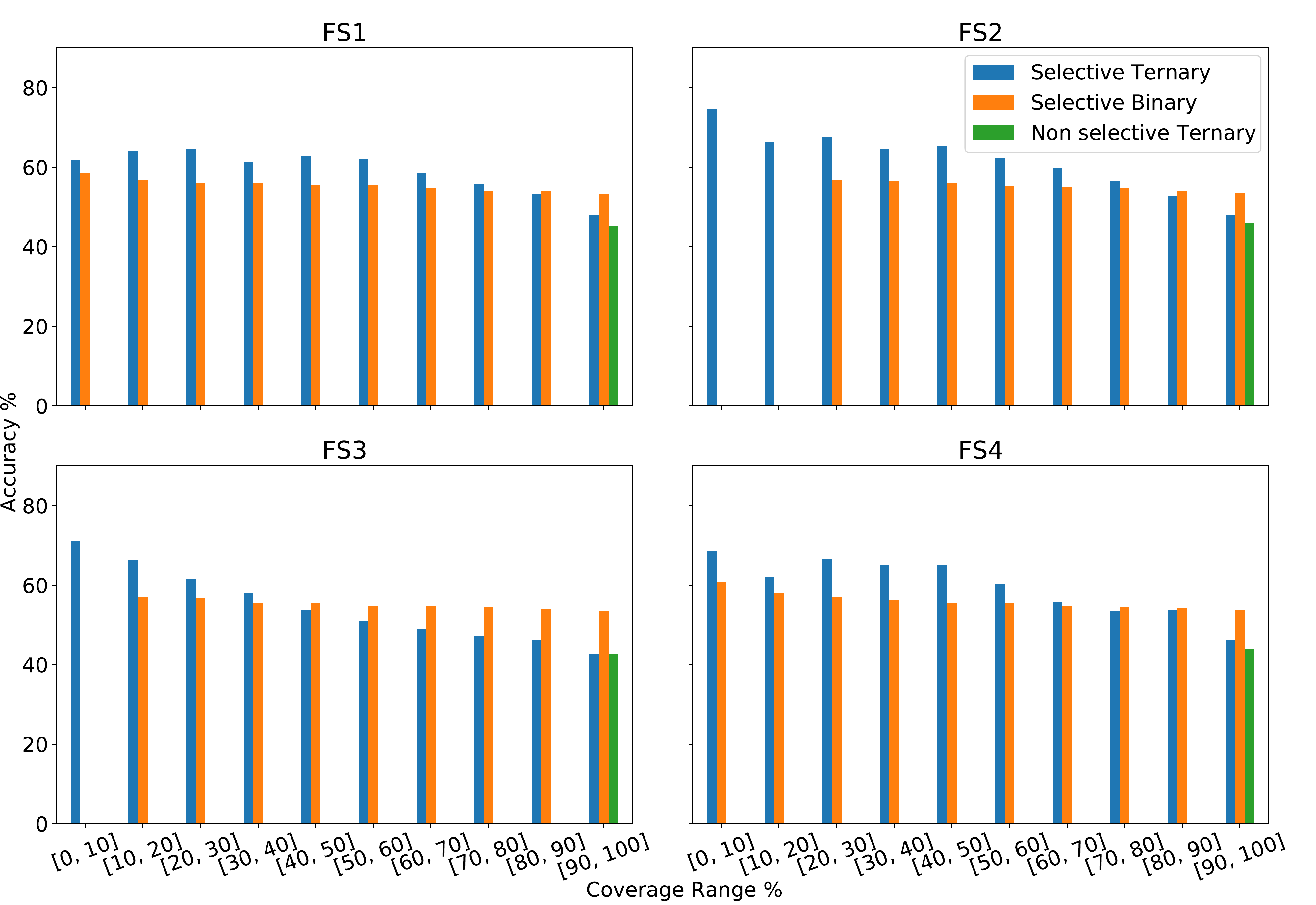}
     }
	 \caption{Accuracy/coverage trade-off of selective binary/ternary LSTM classifiers.}      
	 \label{acc_cov}
\end{figure}

The first takeaway is that selective classification works as expected: By being
selective and reducing coverage we are able to improve our accuracy on those 
data points where we do make a prediction.
This can be seen in Tables~\ref{tab:bin_acc} and~\ref{tab:tern_acc}, which show, for binary
and ternary classifiers respectively, the chosen realized test coverage
rates based on the chosen threshold levels.
The resulting coverage levels show quite stringent selectivity, ranging between
37\% and 63\% for the binary classifiers, and between 17\% and 55\% for ternary
classifiers.
We see that in every single case (i.e., across all classification approaches and 
feature sets), the selectivity results in an improvement in accuracy.
This shows the potential of selective classification; however to properly assess
this potential in the context of designing trading strategies requires backtesting
that we will return to in due course.
 
Figure~\ref{acc_cov} shows the coverage/accuracy trade-off for the selective
classifiers, with results for the ternary non-selective classifier given as a
reference point.
Here we see that, broadly speaking, the accuracy coverage trade-off of the
selective classifiers (binary and ternary) are in line with the results
of~\citep{GeifmanE17}, where for lower coverage values we have higher accuracy
levels.


\begin{table}[!h]
\caption{Binary classifiers: Non-selective, selective accuracy and coverage rates on the test set.}
\label{tab:bin_acc}
\scalebox{0.9}{%

\begin{filecontents*}{csv/all_methods_binary.csv}
LR,FS1,53.24,54.64,63.17
,FS2,53.5,55.79,47.53
,FS3,53.46,55.43,51.32
,FS4,53.47,55.85,47.45
RF,FS1,51.23,52.19,61.14
,FS2,52.13,53.95,56.62
,FS3,52.38,54.2,48.68
,FS4,52.45,54.29,41.86
NN,FS1,52.67,55.24,37.03
,FS2,53.16,55.37,44.53
,FS3,52.5,54.32,44.43
,FS4,52.55,54.2,49.06
LSTM,FS1,53.05,54.95,57.13
,FS2,53.47,55.84,47.77
,FS3,53.28,55.74,37.34
,FS4,53.66,56.52,38.14
\end{filecontents*}
\pgfplotstabletypeset[
	columns/0/.style={
        column name={Models}, string type, column type/.add={}{|},
        assign cell content/.code={
            \ifnum\pgfplotstablerow=0
                \pgfkeyssetvalue{/pgfplots/table/@cell content}{\multirow{4}{*}{##1}}%
            \else
            	\ifnum\pgfplotstablerow=4
	                \pgfkeyssetvalue{/pgfplots/table/@cell content}{\multirow{4}{*}{##1}}%
	             \else
	             	\ifnum\pgfplotstablerow=8
	                	\pgfkeyssetvalue{/pgfplots/table/@cell content}{\multirow{4}{*}{##1}}%
	             	\else
	             		\ifnum\pgfplotstablerow=12
	                		\pgfkeyssetvalue{/pgfplots/table/@cell content}{\multirow{4}{*}{##1}}%
	             		\else
	             			\pgfkeyssetvalue{/pgfplots/table/@cell content}{}%
	             		\fi
	             	\fi
	             \fi
            \fi
        },
    },
col sep=comma,
every head row/.style={before row=\toprule,after row=\midrule},
every nth row={4[-1]}{after row=\midrule},
	every last row/.style={after row=\bottomrule},
 header=false,
 fixed,
 zerofill,
 columns/1/.style={column name=Features, string type, column type/.add={}{|}},
 columns/2/.style={column name={\shortstack{Non-sel.\\ Accuracy \%}},column type={R{1.3cm}}},
 columns/3/.style={column name={\shortstack{Selective\\ Accuracy \%}},column type={R{1.3cm}}},
 columns/4/.style={column name={\shortstack{Coverage\\ \%}},column type={R{1.3cm}}}
 ]{csv/all_methods_binary.csv}
}
\end{table}

The tables also show the relative performance of the different classification
methods, and of the different feature sets.
Ultimately, the results are mixed, but clear cut observations we can make include
the following.
Logistic regression (LR) performed well both for the binary and ternary setups.
Random forests (RF) struggled with the richer feature sets, very possibly due to
overfitting.
LSTMs performed well for both binary and ternary setups, but were clearly beaten
by logistic regression in the ternary setup.
In terms of the feature sets while the results are mixed, it is certainly fair
to say that there is no clear evidence of a benefit to using the richer feature
sets, with FS2 arguably being the best choice on balance. 
 




\begin{table}[!h]
\caption{Ternary classifiers: Non-selective, selective accuracy and coverage rates on the test set,
with the highest in-sample MCC of the buy and sell labels used to choose the multiplier.}
\label{tab:tern_acc}
\scalebox{0.9}{%

\begin{filecontents*}{csv/all_methods_ternary_MCC_DOWNUP.csv}
LR,FS1,44.48,58.61,19.8
,FS2,46.12,63,17.95
,FS3,43.83,59.67,16.83
,FS4,46.25,63,22.34
RF,FS1,54.66,57.63,55.11
,FS2,44.08,49.78,31.25
,FS3,42.71,49.35,24.69
,FS4,42.9,48.71,32.41
NN,FS1,43.16,52.25,30.69
,FS2,44.7,55.85,25.17
,FS3,43.84,49.41,54.48
,FS4,47.27,55.97,47.46
LSTM,FS1,45.29,56.13,33.88
,FS2,45.9,59.07,31.89
,FS3,42.67,55.86,25.61
,FS4,43.92,54.84,24.71
\end{filecontents*}
\pgfplotstabletypeset[
	columns/0/.style={
        column name={Models}, string type, column type/.add={}{|},
        assign cell content/.code={
            \ifnum\pgfplotstablerow=0
                \pgfkeyssetvalue{/pgfplots/table/@cell content}{\multirow{4}{*}{##1}}%
            \else
            	\ifnum\pgfplotstablerow=4
	                \pgfkeyssetvalue{/pgfplots/table/@cell content}{\multirow{4}{*}{##1}}%
	             \else
	             	\ifnum\pgfplotstablerow=8
	                	\pgfkeyssetvalue{/pgfplots/table/@cell content}{\multirow{4}{*}{##1}}%
	             	\else
	             		\ifnum\pgfplotstablerow=12
	                		\pgfkeyssetvalue{/pgfplots/table/@cell content}{\multirow{4}{*}{##1}}%
	             		\else
	             			\pgfkeyssetvalue{/pgfplots/table/@cell content}{}%
	             		\fi
	             	\fi
	             \fi
            \fi
        },
    },
col sep=comma,
every head row/.style={before row=\toprule,after row=\midrule},
every nth row={4[-1]}{after row=\midrule},
	every last row/.style={after row=\bottomrule},
 header=false,
 fixed,
 zerofill,
 columns/1/.style={column name=Features, string type, column type/.add={}{|}},
 columns/2/.style={column name={\shortstack{Non-sel.\\ Accuracy \%}},column type={R{1.3cm}}},
 columns/3/.style={column name={\shortstack{Selective\\ Accuracy \%}},column type={R{1.3cm}}},
 columns/4/.style={column name={\shortstack{Coverage\\ \%}},column type={R{1.3cm}}}
 ]{csv/all_methods_ternary_MCC_DOWNUP.csv}
}
\end{table}
 
\begin{figure}[!h]
  \centering 
  \resizebox{\columnwidth}{!}{
  \includegraphics[width=\textwidth]{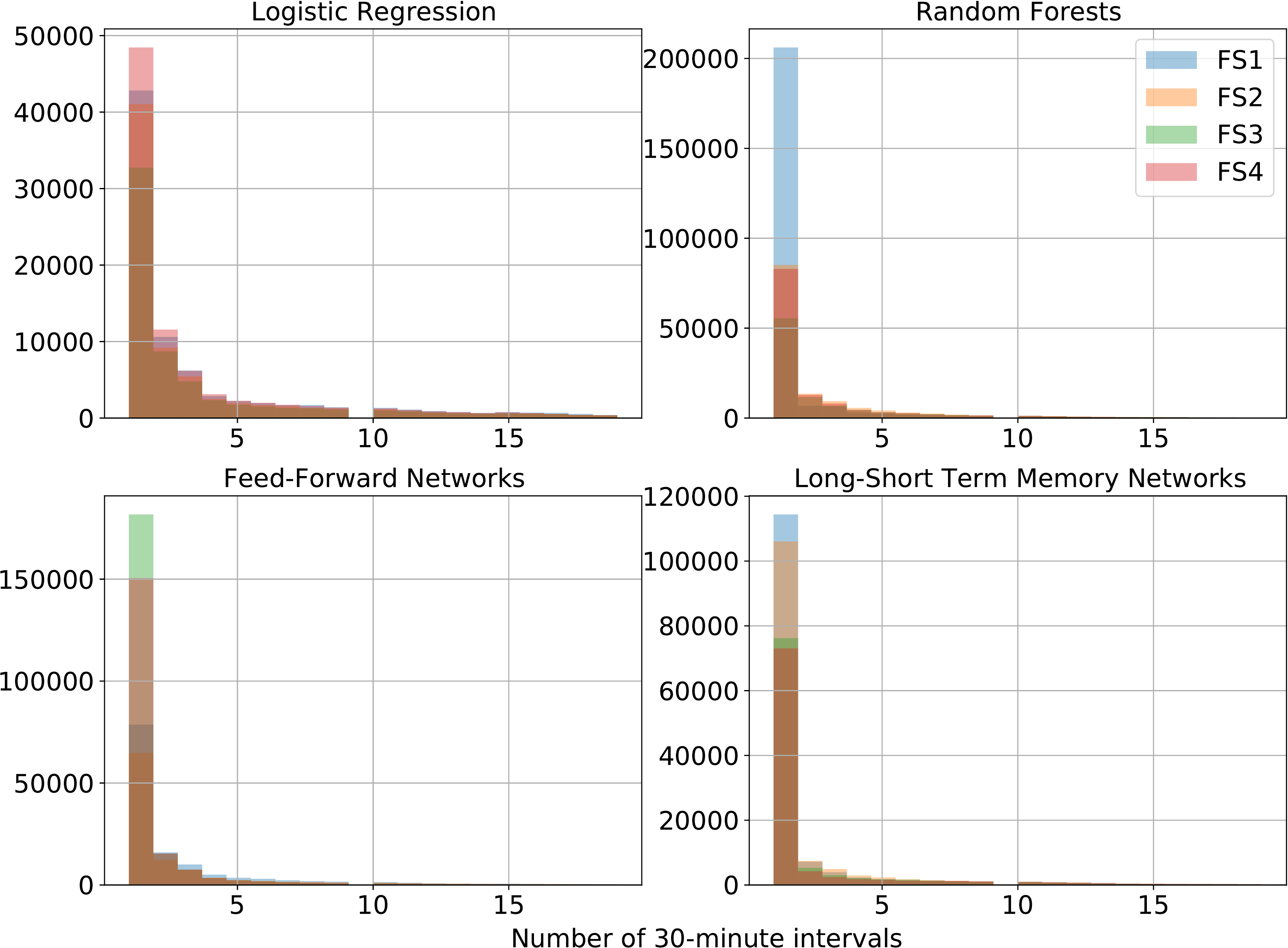}
  } \caption{Time distribution of the non-abstained samples.}
  \label{time_distribution} 
\end{figure}

Given that our selective classifiers abstain on a significant proportion of 
samples, it is natural to explore how the abstentions are distributed through
time. 
For example, do we abstain for very long periods of time?
We investigate this, and, as shown in Figure~\ref{time_distribution}, find that 
this is not the case, namely, that the gaps between predictions are generally
not that large.
Across all classification methods and for all features sets, the distribution of
gaps between predictions appears to follow a power law distribution, and 
the majority of gaps are between $30$-minutes (the smallest possible) and 
$2.5$ hours. 
This is certainly not a requirement for a good classifier (and resulting trading
strategy), but is reassuring in so far as it shows that suitable conditions for
making predictions do occur regularly.

\emph{Binary versus ternary classification.}
To finish this section, we discuss the performance of binary versus ternary 
classification.
Tables~\ref{tab:bin_acc} and~\ref{tab:tern_acc} show significant differences
between the binary and ternary cases in terms of both coverage and accuracy.
One consistent pattern is that total non-selective accuracy is lower
in the ternary case.
This is not a surprise since the ternary classifier has to be strictly more 
discerning to achieve the same level of total accuracy as the binary classifier.
A clear but not totally consistent pattern is that the total selective accuracy 
is higher in the ternary case, and the coverage is less in the ternary case.
A possible explanation is that a selective ternary classifier
is optimized based on \emph{both} the coverage threshold and the multiplier that determines the 
labelling.
The multiplier is chosen to optimize the MCC of the buy and sell labels.
In the ternary selective case we find that, in general, it picks relatively
high multiplier values which gives a relatively large number of 0 labels
-- this can be seen in Table~\ref{tab:true_labels}, which shows the resulting
distribution of labels in the test set (with only FS1 for brevity), where
between 45 and 54\% of the labels are 0 (flat), significantly higher than
$\frac{1}{3}$.
The coverage threshold is set to optimize the MCC across
all (2 or 3) classes, and in the ternary case this optimization step when combined with
the extra multiplier parameter, is giving better selective accuracy via lower coverage
(higher coverage thresholds).
%
%
Table~\ref{tab:true_labels} also shows differences between models in terms of
the distribution of true labels among all samples and just those where the model
abstains; for example, unlike the other models, the RF (random forests) model
has a very large difference, 54\% versus 36\%, in the percentage of flat labels.
This arises due to the RF model being very certain of its flat predictions, so 
abstaining relatively less for this label.
%


%

\begin{table}[!h]
\caption{True labels percentages across all test data, for all
rows (``All'') and just abstained rows (``Abs.'') for the ternary selective classifiers with FS1. All true labels percentages are
rounded.}
\label{tab:true_labels}
\scalebox{0.9}{%
\begin{filecontents*}{csv/true_labels_ternary_merged.csv}
short,All,26,23,28,27
,Abs.,27,32,30,29
flat,All,49,54,45,47
,Abs.,46,36,41,42
long,All,26,23,28,27
,Abs.,27,32,29,29
\end{filecontents*}
\pgfplotstabletypeset[
	columns/0/.style={
        column name={Label}, string type, column type/.add={}{|},
        assign cell content/.code={
            \ifnum\pgfplotstablerow=0
                \pgfkeyssetvalue{/pgfplots/table/@cell content}{\multirow{2}{*}{##1}}%
            \else
            	\ifnum\pgfplotstablerow=2
	                \pgfkeyssetvalue{/pgfplots/table/@cell content}{\multirow{2}{*}{##1}}%
	             \else
	             	\ifnum\pgfplotstablerow=4
	                	\pgfkeyssetvalue{/pgfplots/table/@cell content}{\multirow{2}{*}{##1}}%
	             	\else
	               		\pgfkeyssetvalue{/pgfplots/table/@cell content}{}%
	             	\fi
	             \fi
            \fi
        },
    },
col sep=comma,
every head row/.style={before row=\toprule,after row=\midrule},
every nth row={2[-1]}{after row=\midrule},
	every last row/.style={after row=\bottomrule},
 header=false,
 fixed,
 columns/1/.style={column name=Rows, string type, column type/.add={}{|}},
 columns/2/.style={column name={\shortstack{LR}},column type={R{0.8cm}}},
 columns/3/.style={column name={\shortstack{RF}},column type={R{0.8cm}}},
 columns/4/.style={column name={\shortstack{NN}},column type={R{0.8cm}}},
 columns/5/.style={column name={\shortstack{LSTM}},column type={R{0.8cm}}}
 ]{csv/true_labels_ternary_merged.csv}
}
\end{table}

Given our intended trading application, whether binary or ternary classification
is better cannot easily be determined by (selective) accuracy, not least because
correct and incorrect classifications can correspond to very different profit
and loss amounts for a trading strategy. 
In the next section, we backtest the resulting trading strategies to explore this 
further.


\section{Trading strategies}
\label{sec:trading}

As we just noted, from a trading perspective some misclassifications are more
costly than others.
We next turn our classifiers into trading strategies which we backtest and compare.

\subsection{Backtesting methodology}

We will hold a position that is consistent with our predictions.
That is, whenever we make a prediction of label -1, we will hold a short position,
and when we predict label 1 we will take a long position.
Position sizing is discussed below, along with slippage which will be applied
whenever we trade (which is determined by the desired position sign, namely long/short/flat).

Thus, the behaviour of our binary/ternary selective/non-selective classifiers will 
be as follows:
\begin{itemize}[leftmargin=0.3cm]
\item
A non-selective binary classifier is thus ``always in the market'' (never flat).
 
\item
A selective binary classifier stays flat (i.e., does not take a position) precisely
when the classifier abstains.

\item
A non-selective ternary classifier stays flat when the predicted label is 0.
 
\item
	A selective ternary classifier stays flat \emph{either} when the predicted
	label is 0, or when the classifier abstains.
\end{itemize}

We trade when the desired position sign changes.
The position size for a given commodity (specified as a number of futures
contracts) is set to be inversely proportional to a 5-day moving average of the 
absolute close-on-close move in dollar terms.
This simple scheme is used so that we can reasonably aggregate profit and
loss across the commodities.
To be conservative, slippage was paid on every contract traded.
The amount of slippage was defined as a multiple $\{0, 0.1, \dots,
0.5\}$ of the tick size for the respective commodity.
The number of contracts traded are determined by the difference between the
current position and desired position, so, for example, if the current position
is 6 contracts (long) and the desired position is 2 contracts (short), then the
resulting trade would sell 8 contracts (paying slippage on each).

The backtest applies the classifiers to each commodity independently and trades
accordingly, aggregating the resulting profit and loss across the commodities.
In order to get a risk-adjusted measure, we compute and report a profit-based
variant of the Sharpe Ratio (i.e., one that assumes a constant underlying cost
to each trade, which we consider fine as we only use the resulting numbers for
roughly assessing relative performance).

\begin{figure}[!h]
  \centering
  \resizebox{\columnwidth}{!}{%
  \includegraphics[width=\textwidth]{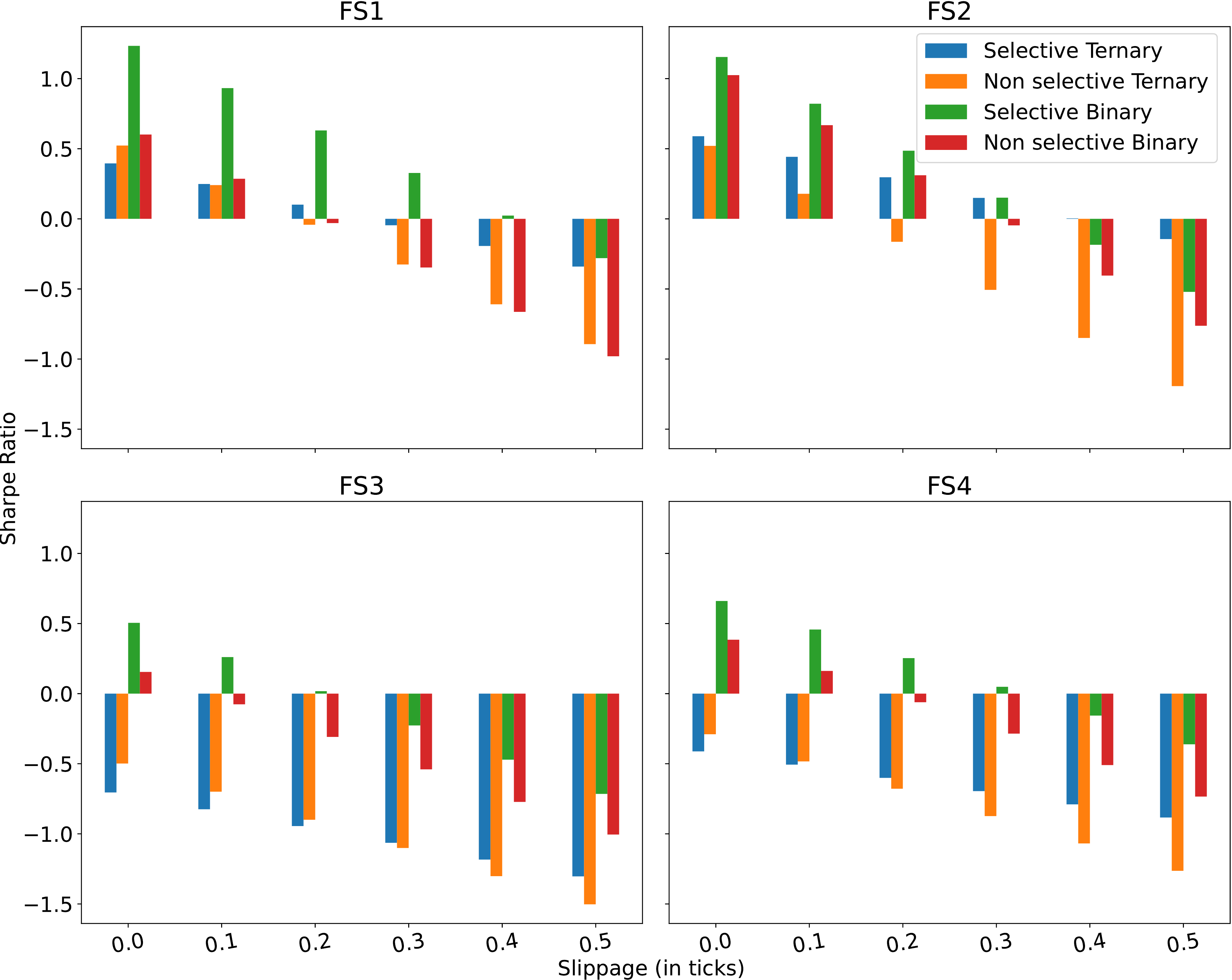}
  }
  \caption{Sharpe Ratios for backtested ternary/ binary selective and
  non-selective Feed-Forward network classifiers.}
  \label{NN_sharpe}
\end{figure}

\begin{figure}[!h]
  \centering
  \resizebox{0.8\columnwidth}{!}{%
  \includegraphics[width=\textwidth]{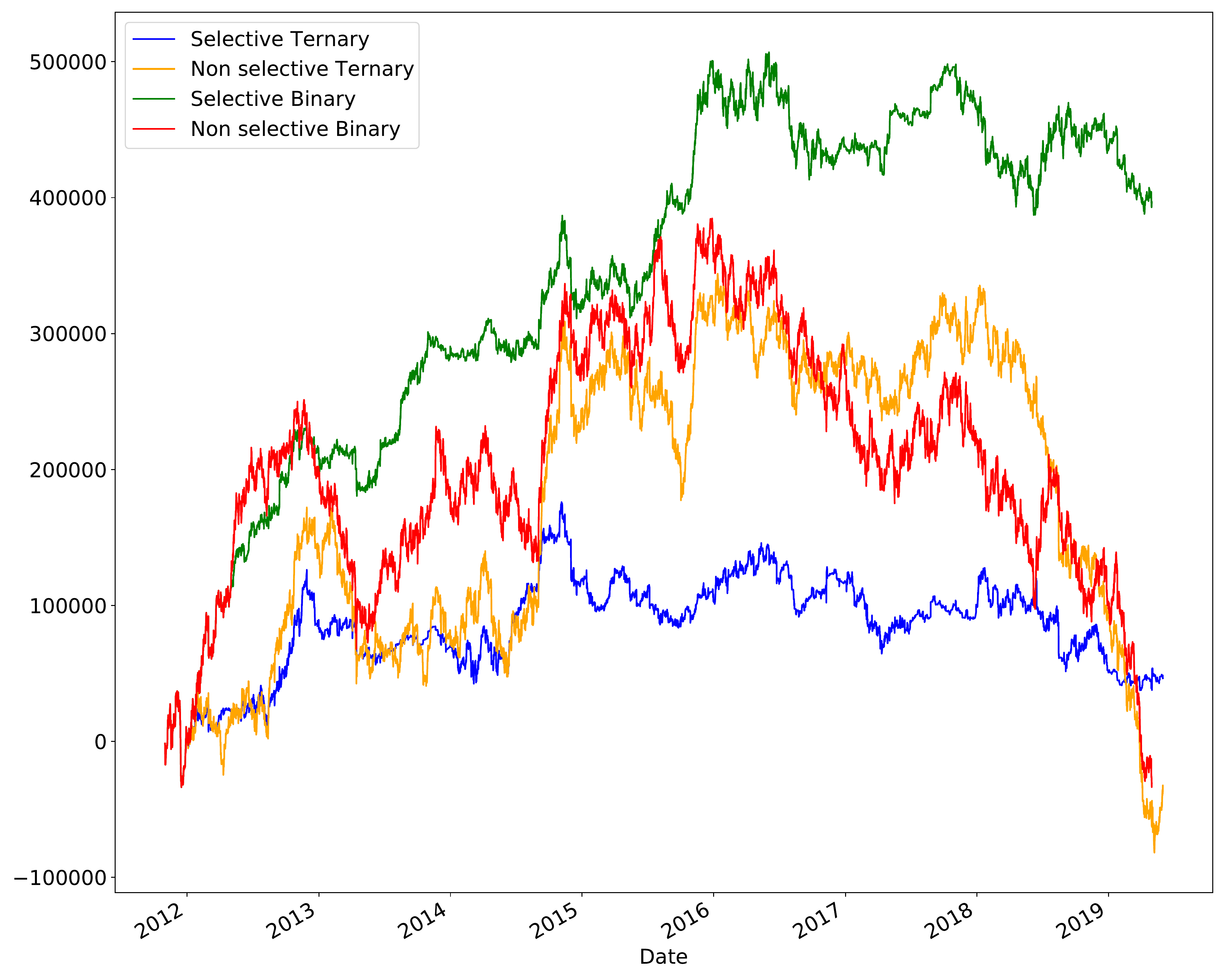}
  }
  \caption{Indicative equity curves corresponding to Figure~\ref{NN_sharpe},
  feature set FS1 and slippage level $0.2$.}
  \label{NN_equity_curve}
\end{figure}

\subsection{Backtesting results}
\label{sec:trading_results}

All reported backtesting results are out-of-sample (see Figure~\ref{WF}).

In Figure~\ref{NN_sharpe}, the results of the feed-forward networks backtesting
process are presented. 
Figure~\ref{NN_equity_curve} shows the equity curves that correspond to 
the FS1 and slippage level 0.2 results within Figure~\ref{NN_sharpe}.
As slippage increases, both binary
and ternary selective classifiers have better Sharpe Ratio values compared to
their respective non-selective classifiers. 
From the four features sets, the FS2
feature set has the best Sharpe Ratio results, taking into account the results
of each classifier for all slippage values. 
The FS3 feature set has the worst
Sharpe Ratio values compared to the other features sets. 
In particular, since FS1 tended to provide better results than FS4, we did not
see the benefits of our richest feature sets\footnote{It is possible that overfitting
was responsible, and where we did $l_2$ regularization, it may be beneficial 
in this regard to try $l_1$ regularization.}. 

In Table~\ref{tab:sharpe}, the Sharpe Ratio results of all the classifiers and
features sets are presented for slippage set to $0.3$ ticks.
At this level of slippage, many configurations are not profitable, but some are,
including:
With all four feature sets, the logistic regression (LR) binary selective classifiers
were profitable, as are some other configurations of this type of classifier;
for FS1 and FS2, the feed-forward network (NN) binary selective classifiers were profitable;
for FS1 and FS2, the random forests (RF) ternary selective classifiers were profitable.
The classifiers using LSTMs are never profitable, which may well just reflect
the difficulty in training these types of models.

\begin{table}[!h]
\caption{Sharpe Ratios for the binary/ternary selective
and non-selective classifiers with slippage at $0.3$ ticks.}
\label{tab:sharpe}
\resizebox{0.9\columnwidth}{!}{%

\begin{filecontents*}{csv/slippage_0.3.csv}
LR,FS1,0.0221,0.1637,-0.1211,-0.5977
,FS2,0.213,0.0187,0.032,-0.9724
,FS3,0.1466,-0.0795,-0.1571,-0.8217
,FS4,0.2842,-0.0345,-0.2022,-0.3057
RF,FS1,-1.6109,-1.6004,-0.3523,-1.4589
,FS2,-0.5705,-0.6137,-0.2385,-1.0425
,FS3,-0.0273,-0.6097,0.5744,-1.0272
,FS4,-0.409,-0.7546,0.3482,-0.104
NN,FS1,0.3272,-0.3469,-0.0458,-0.3256
,FS2,0.1508,-0.0466,0.1498,-0.5066
,FS3,-0.227,-0.5401,-1.064,-1.1005
,FS4,0.0486,-0.28528,-0.6951,-0.8732
LSTM,FS1,-0.8638,-0.8895,-0.2321,-0.7596
,FS2,-0.8073,-0.6702,-0.2907,-1.225
,FS3,-0.7045,-0.872,-0.4,-1.2155
,FS4,-0.4279,-0.6941,-0.1198,-1.1015
\end{filecontents*}
\pgfplotstabletypeset[
	columns/0/.style={
        column name={Models}, string type, column type/.add={}{|},
        assign cell content/.code={
            \ifnum\pgfplotstablerow=0
                \pgfkeyssetvalue{/pgfplots/table/@cell content}{\multirow{4}{*}{##1}}%
            \else
            	\ifnum\pgfplotstablerow=4
	                \pgfkeyssetvalue{/pgfplots/table/@cell content}{\multirow{4}{*}{##1}}%
	             \else
	             	\ifnum\pgfplotstablerow=8
	                	\pgfkeyssetvalue{/pgfplots/table/@cell content}{\multirow{4}{*}{##1}}%
	             	\else
	             		\ifnum\pgfplotstablerow=12
	                		\pgfkeyssetvalue{/pgfplots/table/@cell content}{\multirow{4}{*}{##1}}%
	             		\else
	             			\pgfkeyssetvalue{/pgfplots/table/@cell content}{}%
	             		\fi
	             	\fi
	             \fi
            \fi
        },
    },
 col sep=comma,
 every head row/.style={before row=\toprule,after row=\midrule},
 every nth row={4[-1]}{after row=\midrule},
 every last row/.style={after row=\bottomrule},
 header=false,
 fixed,
 zerofill,
 columns/1/.style={column name=Features, string type, column type/.add={}{|}},
 columns/2/.style={column name={\shortstack{Selective\\ Binary}},column type={R{1.2cm}}},
 columns/3/.style={column name={\shortstack{Non-sel.\\ Binary}},column type={R{1.3cm}}},
 columns/4/.style={column name={\shortstack{Selective\\ Ternary}},column type={R{1.3cm}}},
 columns/5/.style={column name={\shortstack{Non-sel.\\ Ternary}},column type={R{1.3cm}}}
 ]{csv/slippage_0.3.csv}
}
\end{table}

We consider these results to be a promising proof of concept for this trading 
approach, especially given that 
many parameters of the method were not optimized and just set as 
intuitively reasonable choices.
This also applies to the feature sets; there is a lot of scope to
improve these and other aspects of the setup.

\section{Conclusions/ Future Work}
This study presents an application of selective classification in futures time
series, and backtests trading strategies based on the classifier's predictions.
We found that:
\begin{itemize}[leftmargin=0.3cm]
\itemsep2mm
	\item The selective classifiers performed better than 
	their non-selective counterparts in terms of accuracy.
	\item Lower coverage values resulted in higher accuracy
	levels.
	\item The selective classifiers had better backtesting
	results compared to their respective non-selective
	classifiers.
	\item The results did not demonstrate any advantage of the richer
	feature sets.
	\item Selective classification reduced the misclassification
	percentages by abstention, which helped the trading 
	strategies to avoid losses.
	\item The selective binary classifiers had better 
	backtesting results compared to the selective and non-selective
	ternary classifiers. 
\end{itemize}
The results show the potential of selective classification, as the selective
classifiers performed better than their non-selective counterparts in terms of
accuracy and backtesting results. 
It would be interesting to explore the use of other selective
classification algorithms, such as~\cite{GeifmanE19,ThulasidasanBBC19}.
As discussed at the end of Section~\ref{sec:classresults}, the ternary 
classifiers worked quite differently from the binary classifiers and appeared
promising when looking only at accuracy, but in
the end provided worse backtesting results. 
It would be interesting to explore hyperparameter optimization based on 
backtesting results directly, rather than the MCC criterion.

\begin{acks}
The authors would like to acknowledge the support of this 
work through the EPSRC and ESRC Centre for Doctoral Training 
on Quantification and Management of Risk  Uncertainty in 
Complex Systems Environments Grant No. (EP/L015927/1).
\end{acks}

\bibliographystyle{ACM-Reference-Format}
\bibliography{papers}

\end{document}